\documentclass[proof]{WileyASNA-v1}

\articletype{Article Type}%

\received{6 December 2018}

\begin{document}

\title{High Energy Processes in Wolf-Rayet Stars}

\author[1]{Stephen L. Skinner*}

\author[2]{Werner Schmutz}

\author[3]{Manuel G\"{u}del}

\author[4]{Svetozar Zhekov}


\address[1]{\orgdiv{CASA}, \orgname{Univ. of Colorado}, \orgaddress{\state{Boulder, CO 80309-0389}, \country{USA}}}

\address[2]{\orgdiv{PMOD}, \orgname{WRC}, \orgaddress{\state{Davos Dorf}, \country{Switzerland}}}

\address[3]{\orgdiv{Dept. of Astrophysics}, \orgname{Univ. of Vienna}, \orgaddress{\state{Vienna}, \country{Austria}}}

\address[4]{\orgdiv{Inst. of Astron. \& Natl. Astron. Obs.}, \orgname{Bulgarian Acad. Sci.}, \orgaddress{\state{Sofia}, \country{Bulgaria}}}

\corres{*\email{stephen.skinner@colorado.edu}}


\abstract{Wolf-Rayet (WR) stars are massive  ($\geq$10 M$_{\odot}$) 
evolved stars  undergoing advanced 
nuclear burning in their cores, rapidly approaching the end of their lives 
as supernovae. Their powerful winds enrich the interstellar medium with 
heavy elements, providing raw material for future generations of stars.
We briefly summarize high-energy processes in WR stars, focusing
mainly on their X-ray emission. We present new results from  
{\it Chandra} observations of the eclipsing WR$+$O binary CQ Cep
 covering a full orbit which stringently test X-ray emission models.
}




\maketitle


\section{Wolf-Rayet Stars: Overview}\label{sec1}

  WR stars are broadly classified into three subtypes based on
  their optical spectra: nitrogen-type (WN), carbon-type (WC), and
  oxygen-type (WO). WO stars are the most evolved and only a
  few are known in the Galaxy. WR stars have high effective
  temperatures $>$20,000 K and tremendous winds with typical
  mass-loss rates $\dot{M}$ $\sim$10$^{-5}$ M$_{\odot}$ yr$^{-1}$.
  Terminal wind speeds for WN and WC stars are typically
  V$_{\infty}$ $\sim$1000 - 2500 km s$^{-1}$, and even higher for WO stars.
  Most, but not all,  WR stars are strong  X-ray sources. Their X-ray 
  emission is usually attributed to shocks associated with 
  their supersonic winds, but shock model predictions are in some cases
  not compatible with the observed X-ray properties. 
  Nonthermal (synchrotron)
  radio continuum emission has been detected in some WR binaries.
  WR binaries are potential sources of $\gamma$-ray
  emission and  sites for Galactic cosmic ray acceleration.

\section{X-Rays}\label{sec2}

\subsection{Single WR Stars}

X-rays have been detected from putatively single (non-binary)
WN and WO stars, but surprisingly not from WC stars.
X-ray luminosities of WN stars span a range of about two orders
of magnitude with typical values log L$_{x}$ $\sim$ 10$^{32\pm1}$ 
ergs s$^{-1}$ (Skinner et al. 2012). The only galactic WO star 
detected so far in X-rays is WR 142 with  log L$_{x}$ $\approx$ 31.3
ers s$^{-1}$ at its {\it GAIA} DR2 distance of 1.74 kpc
(Sokal et al. 2010; Skinner et al. 2018).
WC stars are X-ray faint or X-ray quiet for reasons not yet known,
but strong wind absorption may be partially responsible.
The most stringent upper limit obtained for a single WC
star so far is from  a  {\it Chandra} observation of 
WR 135 (WC8) which gives  log L$_{x}$ $\leq$ 29.99
ergs s$^{-1}$ at its {\it GAIA DR2} distance of 2.11 kpc
(Skinner et al. 2006).

The X-ray spectra of single WN stars can be
acceptably modeled as a two-temperature (2T) optically-thin 
plasma with a cool component
at kT$_{1}$ $\approx$ 0.3 - 0.7 keV [T$_{1}$ $\approx$ 3 - 8 MK]
and a hotter component at 
kT$_{2}$ $\approx$ 2 - 5 keV [T$_{2}$ $\approx$ 20 - 60 MK]
 (Skinner et al. 2010; 2012). Such 2T models do not fully
 reflect the true physical conditions since the X-ray plasma
 is distributed over a range of temperatures.

The temperature of the cool component is consistent
with predictions of radiative wind shock models, 
which posit X-ray production from shocks that form
in the wind as a result of line-driven instabilities
(e.g. Lucy \& White 1980).
Such models have had success in explaining
the soft X-ray emission of some O-type stars, but their relevance
to WR stars with much higher mass-loss rates and wind
speeds remains to be determined. But the hotter plasma,
which is prominent in single  WN star spectra
and in the spectrum of the WO star WR 142, is not
anticipated from radiative shock models. Its
origin is so far unexplained.

\subsection{WR Binaries}

Massive WR binaries are typically strong
X-ray sources. This includes WC systems 
such as $\gamma^2$ Vel (=  WR 11; WC8$+$O7) and
WN systems such as WR 147 (WN8$+$B0.5V). 
In some cases,  the X-ray
emission is bright enough  to obtain high-resolution 
X-ray grating spectra, allowing individual emission lines
to be identified and studied 
(e.g.  $\gamma^2$ Vel, Skinner et al. 2001; 
WR 140, Pollock et al. 2005; WR48a, Zhekov et al. 2014). 
Line information, when 
available,  constrains  the plasma
temperature distribution, metal abundances, and
distance from the star(s) where the line forms.

The X-ray emission of WR binaries is potentially an admixture
of multiple components including that of 
the individual stars (and their winds) and  colliding
wind (CW) shock emission originating between the stars,
or near the surface of the star with the 
weaker wind (Usov 1992).
In most cases, these different components cannot be
spatially-resolved with  current generation 
X-ray telescopes.

The maximum ~CW shock temperature for an adiabatic shock is 
kT$_{cw}$ $\approx$ 1.96$\mu$[V$_{\perp}$/1000 km/s]$^2$ keV.
Here, $\mu$ is the mean atomic weight (amu) in the wind 
($\mu$ $\approx$ 4/3 for He-dominated WN winds) and
V$_{\perp}$ is the wind velocity component perpendicular
to the shock front. The hottest plasma is predicted to lie on
or near the line-of-centers where V$_{\perp}$ $\approx$
V$_{\infty}$, if the winds have reached terminal speeds. 
This latter condition will be satisfied in wide
binaries (P$_{orb}$ $\sim$ years) but not in close 
binaries (P$_{orb}$ $\sim$ days). For typical WR 
wind speeds V$_{\infty}$ $\approx$ 1000 - 2500 km/s, 
maximum shock temperatures 
kT$_{cw,max}$ $\approx$ 2 - 12 keV are expected.
Temperatures  in this range are indeed observed in 
some WR binaries, but also in some (apparently)
single WR stars.

\section{The Eclipsing WR Binary CQ Cep}\label{sec3}

CQ Cep (= WR 155) is an eclipsing WN6$+$O9 binary system 
in  a near-circular high inclination 1.64 d orbit (Demircan et al. 1997). 
The masses and radii of the two stars are nearly equal 
(M$_{*}$ $\approx$ 21 M$_{\odot}$, R$_{*}$ $\approx$ 8 R$_{\odot}$). 
Their separation is
$D$  $\approx$  20 R$_{\odot}$, placing the two stars 
nearly in contact (Demircan et al. 1997). At such
close separation, the winds  will not have  reached terminal
speeds before colliding. The higher momentum  of the
WR wind will overpower the O star wind and 
the CW shock will form at or near the O star surface. 
CQ Cep is a superb system for testing CW model predictions
at close spacing where the winds
will be at subterminal speeds and radiative 
cooling may be important (Stevens et al. 1992).

We have observed CQ Cep with the  {\it Chandra} X-ray Observatory
using the ACIS-S CCD imaging spectrometer over a full
orbit (Table 1). The first half of the orbit was observed in a single 
uninterrupted observation in March 2013,  capturing
the O star passing in front of the WR star at $\phi$ = 0 (Skinner et al. 2015). 
The second half was observed in four exposures during Feb.-March 2017,
again capturing the O star in front as well as the WR star in front. 
Simultaneous optical light curves were obtained using 
the {\it Chandra} Aspect Camera Assembly (ACA).
The main objective was to search for X-ray variability
during eclipses, which is expected if the hottest X-ray
plasma is confined to the region on or near the 
line-of-centers between the two stars.

Both the primary and secondary
visual eclipses are clearly seen in the
optical light curves (Fig. 1-top panels). Comparison of
the  overlapping ACA light curves indicates that the 
system was slightly brighter in 2013 than in 2017.
Optical variability is also apparent from
significant scatter in the high-precision {\it GAIA} photometry
at similar phases but different epochs. The times of 
minimum ACA optical  brightness at $\phi$=0 in 2013 and
2017 give an accurate orbital period P = 1.641239 ($\pm$8.0e$-$07) d.
Analysis of historical data suggests that the period
may be variable (Koenigsberger et al. 2017).

\begin{center}
\begin{table}[t]%
\centering
\caption{Chandra Observations of CQ Cep\label{tab1}}%
\tabcolsep=0pt%
\begin{tabular*}{20pc}{@{\extracolsep\fill}lccccc@{\extracolsep\fill}}
\toprule
\textbf{ObsId} & \textbf{Start Date}  & \textbf{Duration}   & \textbf{Phase}          & \textbf{Rate}        & \textbf{P$_{var}$}   \\
               &                      &   (ks)              & ($\phi$)                & (c ks$^{-1}$)        &                      \\
\midrule
14538 & 2013 Mar 19  & 85.6  & ($-$0.10, 0.50)   & 24.7$\pm$5.6\tnote{$\dagger}$  & 0.22 \\
17734 & 2017 Feb 27  & 18.3  & ($-$0.07, 0.06)   & 20.9$\pm$4.0                              & 0.16 \\
20016 & 2017 Mar  4   & 16.4  & (0.66, 0.78)        & 20.3$\pm$3.9                              & 0.07 \\
20017 & 2017 Mar 5    &  33.9 & (0.43, 0.67)        & 20.5$\pm$4.1                              & 0.05 \\
20018 & 2017 Mar 5    &  19.2 & (0.79, 0.93)        & 20.6$\pm$4.7                              & 0.04 \\
\bottomrule
\end{tabular*}
\begin{tablenotes}
\item O star is in front at $\phi$ = 0.
P$_{var}$ is the probability of variable count rate based on
the Chandra {\it glvary} statistical test.
\item[$\dagger$] ACIS-S effective area was higher in 2013.
\end{tablenotes}
\end{table}
\end{center}

\begin{figure}[h]
\centerline{\includegraphics[width=35mm,height=70mm,angle=-90]{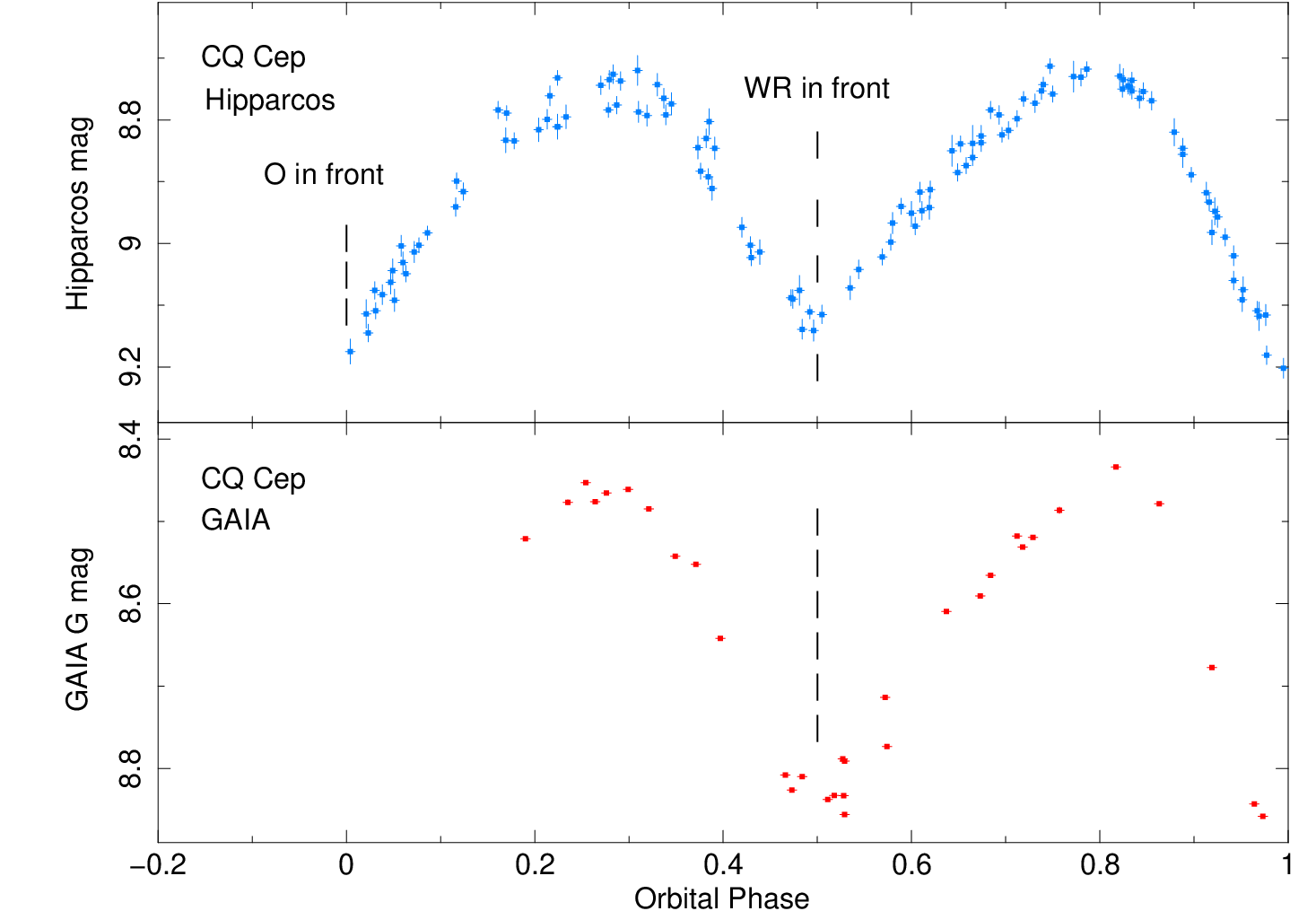}}
\centerline{\includegraphics[width=35mm,height=70mm,angle=-90]{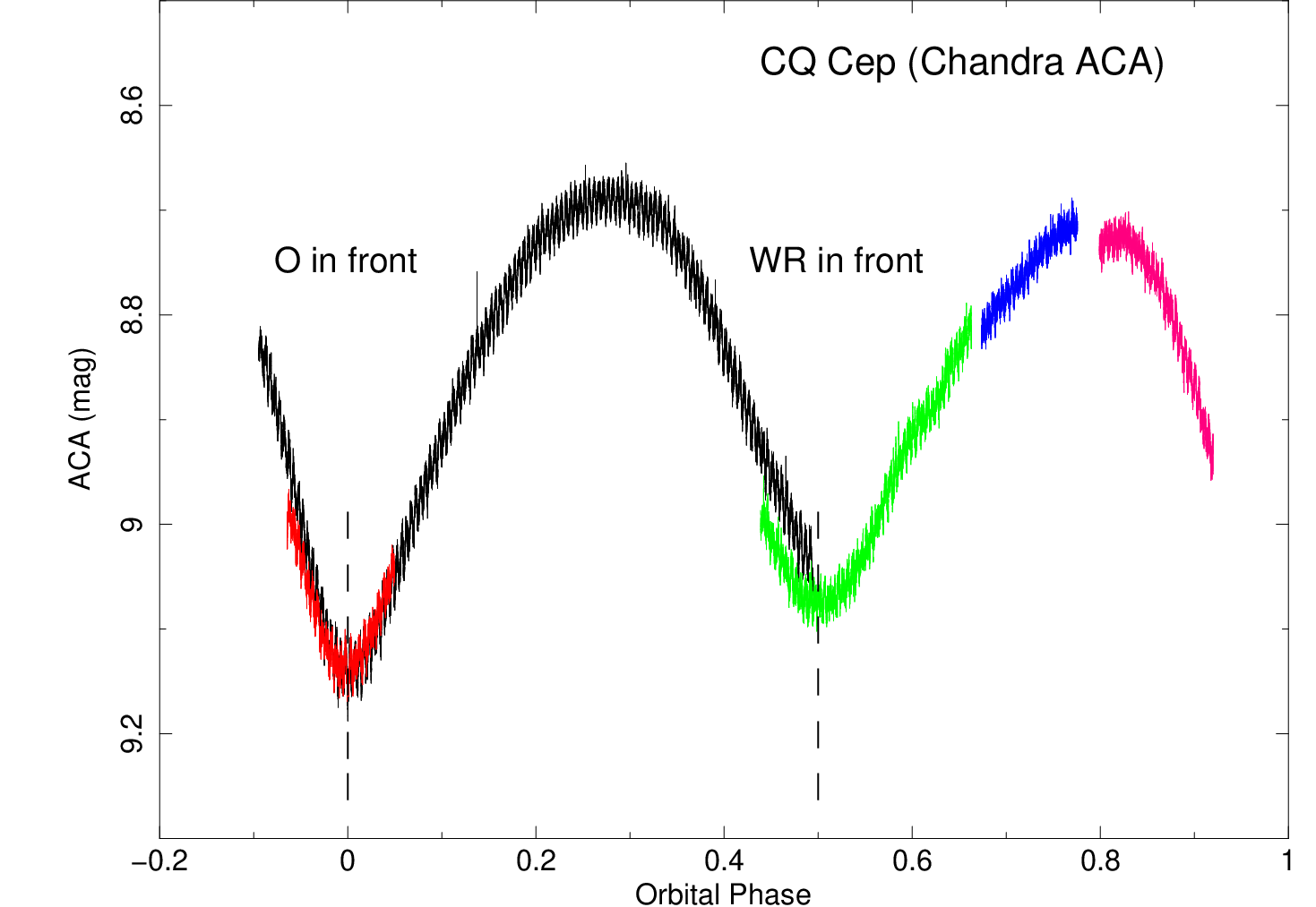}}
\centerline{\includegraphics[width=35mm,height=70mm,angle=-90]{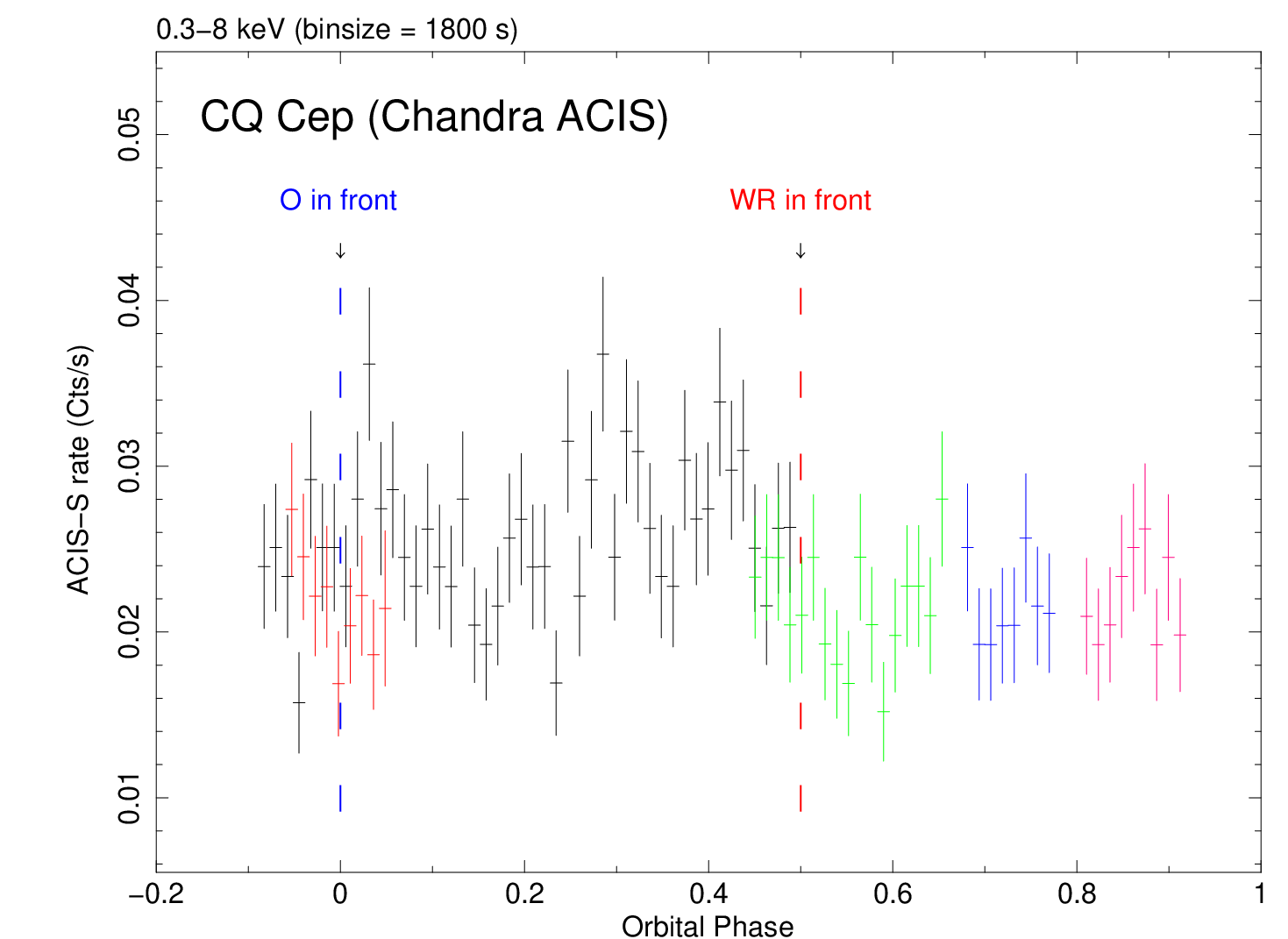}} 
\centerline{\includegraphics[width=35mm,height=70mm,angle=-90]{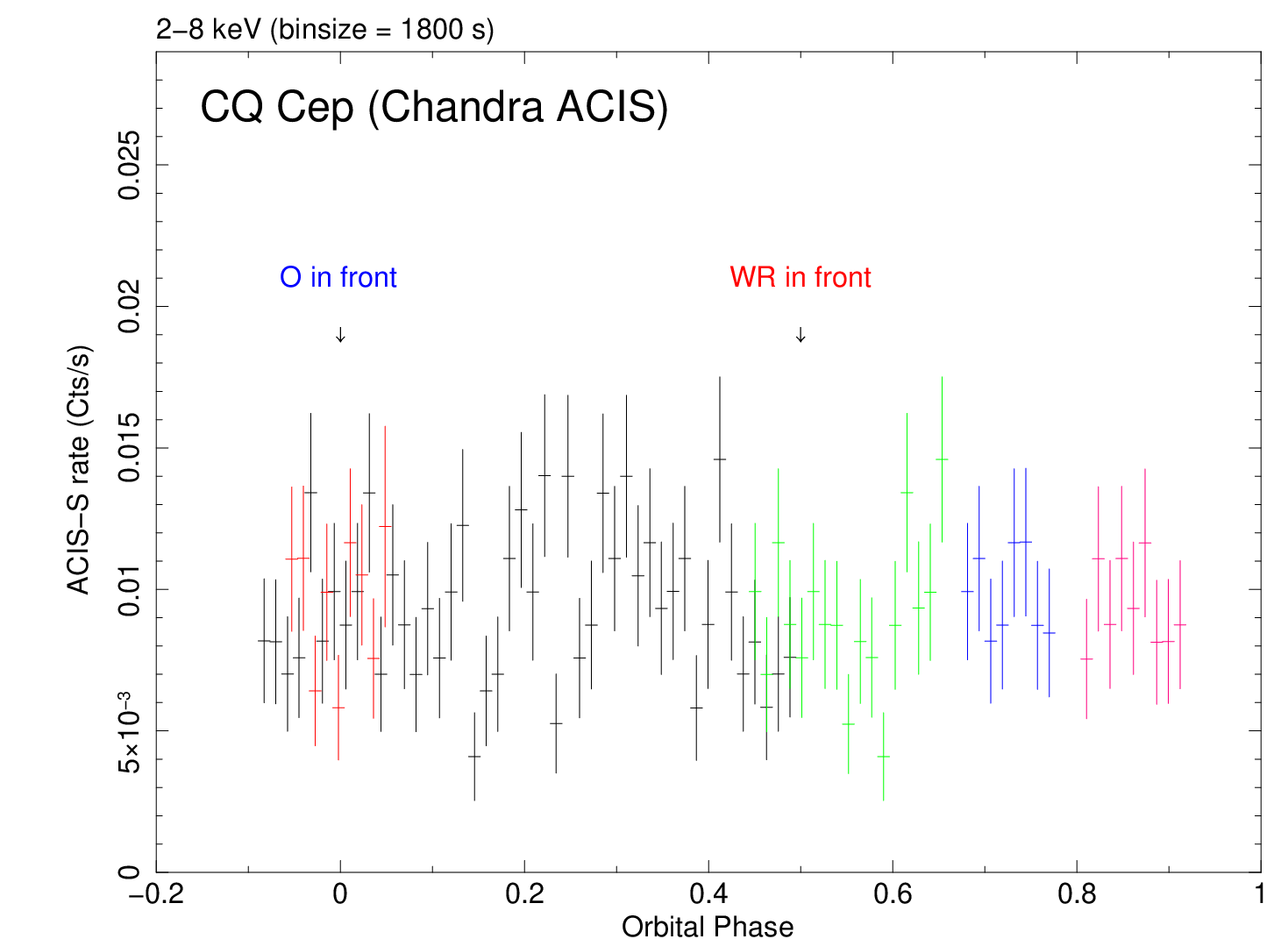}} 
\caption{Phased CQ Cep light curves. 
Top: Optical (Hipparcos epoch 1990.0-1993.1, 
GAIA $G$ mag epoch 2013.3-2014.6; 1$\sigma$ errors). 
2nd row:  Optical (Chandra  ACA), color coded by
observation date (Table 1).
3rd row: Chandra X-ray broad (0.3-8 keV).
 ~Bottom:  Chandra X-ray hard (2-8 keV).\label{fig1}}
\end{figure}

In contrast, the phased X-ray light  curves (Fig. 1-bottom panels)
show only low-level fluctuations. No statistically significant
variability is present and the  mean count rates of the  observations 
agree to within $\pm$1$\sigma$ (Table 1). ACIS-S effective area declined
during 2013-2017 due to contaminant buildup, resulting in lower
count rates in 2017.

Phase-resolved X-ray spectra obtained in 2017 are shown in Figure 2-top.
All four spectra are  similar and  spectral fits show no significant
difference in  their observed X-ray fluxes.  The 
median energies of the photons detected in the observations
are nearly identical, i.e.  E$_{50}$ = 1.88, 1.90, 1.87, and 1.87 keV
(background is negligible). The spectra  are
overlaid in Figure 2-bottom. Several blended 
emission lines are  detected  including Ne X, Mg XI, 
Si XIII, S XV, Ar XVII and possibly
Ne IX and  Ca XIX. Faint Fe K (Fe XXV) emission 
was detected in the 2013 observation (Skinner et al.  2015),
tracing very hot plasma at T $\sim$ 63 MK.

Spectral fits with non-solar abundance 2T models 
are able to reproduce the spectra. 
The inferred absorption (N$_{\rm H}$) and plasma 
temperatures (kT) are somewhat model dependent,
as is the intrinsic (unabsorbed) X-ray luminosity (L$_{x}$).
Fits of the 2013 spectrum were discussed in Skinner et al. (2015).
Table 2 compares fits of the 2013 and 2017 spectra using a
a 2T plane-parallel shock model. The temperature of the hot component 
(kT$_{2}$ $\approx$ 3.4-3.6  keV) is about twice the predicted value
for the WN6 wind shocking onto the surface of
the O star companion at subterminal speed (Skinner et al. 2015).
However, the predicted shock temperature depends on the assumed
WR wind acceleration profile in the region between the stars, 
which is not well-constrained observationally.

The intrinsic (unabsorbed) X-ray luminosity of CQ Cep
at its {\it GAIA} distance (Table 2) is  at the high end of 
the range for WR stars. The predicted luminosity for a WR
wind shocking onto the surface of an O-type companion
of radius $R$ at separation $D$ is
L$_{x,cw}$ = 0.125($R/D$)$^{2}$L$_{wind,wr}$$F$.
The WR wind kinetic luminosity is
L$_{wind,wr}$ = (1/2)$\dot{M}_{wr}$V$^{2}_{\infty,wr}$.
The correction factor $F$ $<$ 1  accounts for
changes in the direction and magnitude of the WR wind 
velocity vector across the O-star surface, which must
be considered in closely-spaced  binaries where the 
stellar radii are nearly equal, as in CQ Cep.
Inserting appropriate values for CQ Cep,
one obtains log L$_{x,cw}$ = 35.11 - 35.27 ergs s$^{-1}$
(Skinner et al. 2015), which is about fifty times
greater than observed. Similar discrepancies have been
noted in other close WR binaries and  explanations
involving inhomogeneous winds have been proposed (Zhekov 2012).

The absence of significant X-ray variability during eclipses 
is remarkable in comparison to the dramatic optical variability.
This lack of variability, even in the high  2-8 keV energy range,
 indicates that the hottest X-ray plasma occupies a region 
 much larger than the scale of the binary system and is not
 localized near the line-of-centers. 
If the hot plasma does  form in a CW shock, then
orbital motion may have distorted the shape of the 
interaction region and displaced it away from 
the line-of-centers (Skinner et al. 2015, and
references therein).

\begin{figure}[t]
\centerline{\includegraphics[width=52mm,height=70mm,angle=-90]{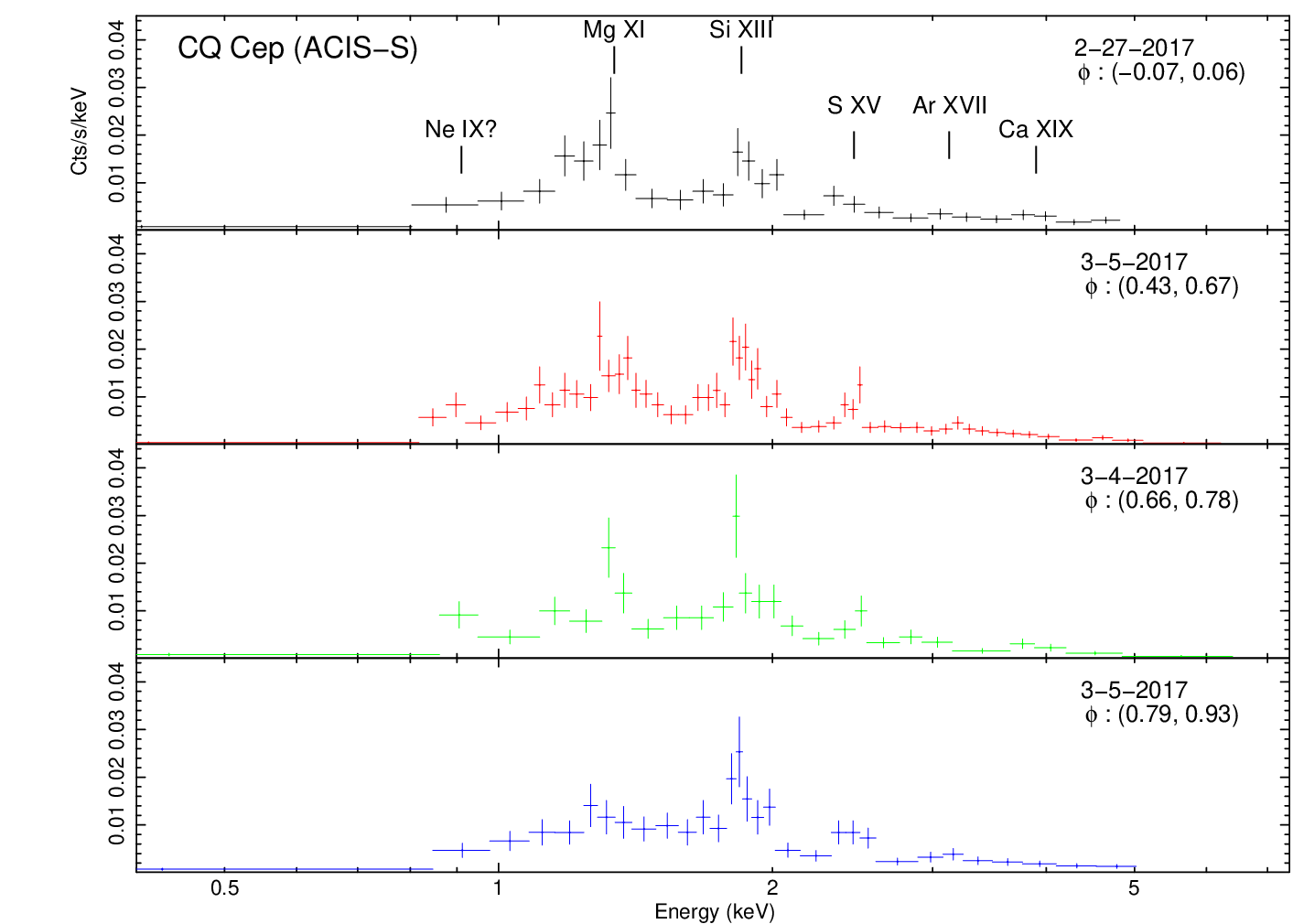}}
\vspace{0.2cm}
\centerline{\includegraphics[width=35mm,height=70mm,angle=-90]{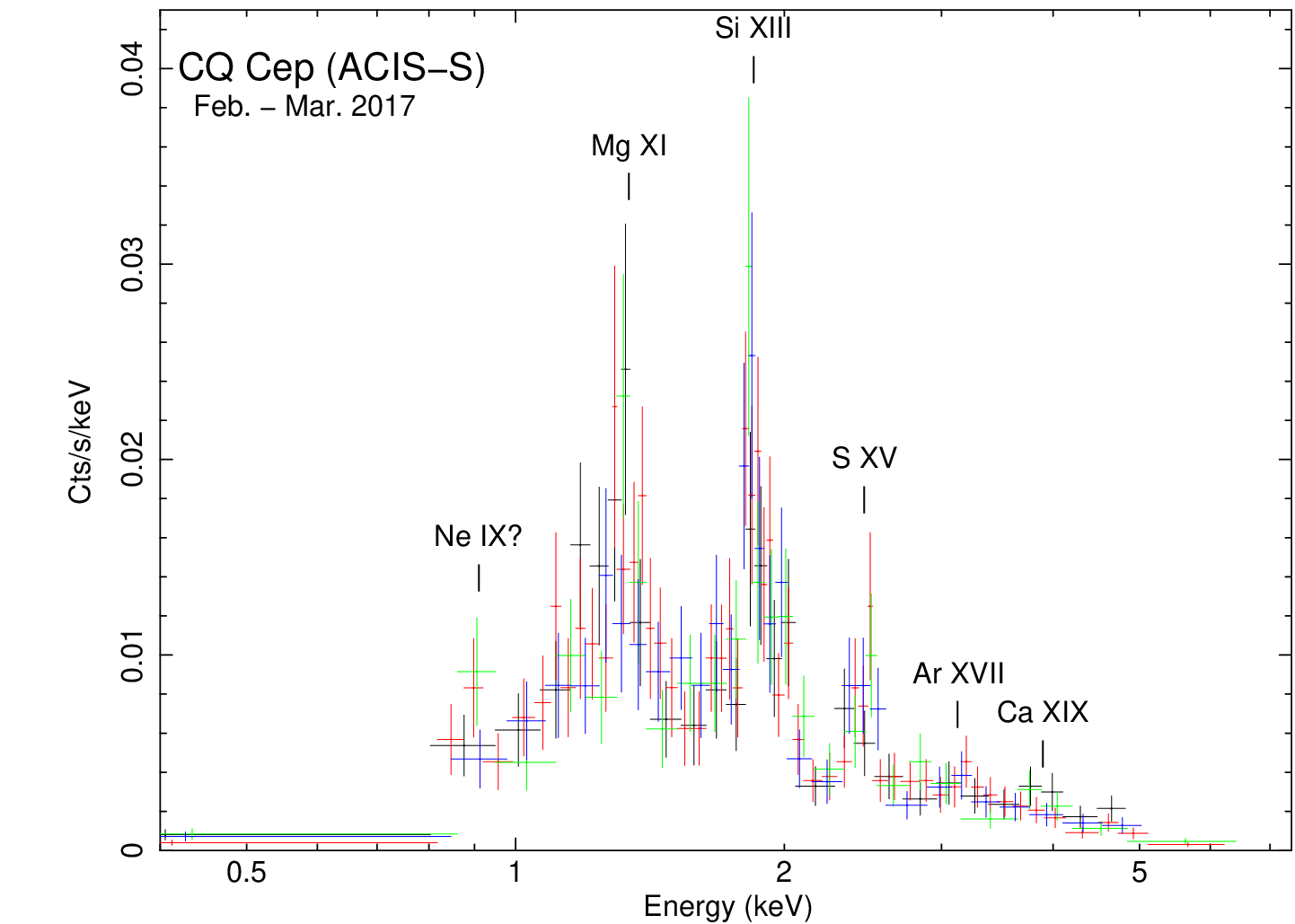}}
\caption{Chandra X-ray spectra of CQ Cep obtained in 2017.
Top: Separated by observation (Table 1).~
Bottom:  All four spectra overlaid.\label{fig2}}
\end{figure}

\begin{center}
\begin{table}%
\centering
\caption{CQ Cep Spectral Fits (shock model)\label{tab2}}
                                                                                                                                                                                               
\tabcolsep=0pt%
\begin{tabular*}{20pc}{@{\extracolsep\fill}lcccc@{\extracolsep\fill}}
\toprule
\textbf{Year}     &    \textbf{N$_{\rm H,1}$, N$_{\rm H,2}$}  & \textbf{kT$_{1}$, kT$_{2}$} & \textbf{F$_{x}$\tnote{$\dagger$}} & \textbf{log L$_{x}$\tnote{$\ddagger$}}   \\
                &   (10$^{21}$ cm$^{-2}$)                   & (keV)                       &                                   & (erg s$^{-1}$)   \\
\midrule
2013            & 4.4[1.7], 0.3[0.1]     & 0.6[0.1], 3.4[0.6]   & 2.09   & 33.43 \\
2017            & 6.5[2.7], 0.3[0.1]     & 1.0[0.3], 3.6[0.9]   & 2.20   & 33.53 \\

\bottomrule
\end{tabular*}
\begin{tablenotes}
\item Fits are based on a 2T plane-parallel shock model (XSPEC {\it vpshock}).
The 2017 values are from a simultaneous fit of all 4 spectra.
The absorption N$_{\rm H}$ of each temperature component was allowed
to vary independently. The tabulated N$_{\rm H}$ values do not include interstellar absorption,
held fixed at N$_{\rm H}$ = 4.4e21 cm$^{-2}$.
Solar abundances were used for the cool component
and generic WN abundances for the hot component (Skinner et al. 2015).
Brackets enclose 1$\sigma$ errors.
\item[$\dagger$] Absorbed flux (0.3-8 keV).
  Units: 10$^{-13}$ ergs cm$^{-2}$ s$^{-1}$.
\item[$\ddagger$]Unabsorbed X-ray luminosity (d=3.32 kpc; GAIA DR2).
\end{tablenotes}
\end{table}
\end{center}

\section{Nonthermal Radio Emission}

Thermal (free-free) radio continuum emission is produced 
by the strong  photoionized winds of WR stars.  In addition,
radio continuum emission characterized by negative or flat
spectral indices $\alpha$ $\leq$ 0 
(flux density varies with frequency as F$_{\nu}$ $\propto$ $\nu^{\alpha}$)
interpreted as synchrotron emission
has been detected in several  WR$+$OB binaries (Abbott et al. 1986).
It is believed that the nonthermal radio emission 
is produced in the CW region between the stars where
electrons are accelerated to relativistic energies by
shocks, but magnetic fields may also play a 
role (Kissmann et al. 2016). Support for the
origin of nonthermal radio emission in the CW region
comes from spatially-resolved radio observations of
WR 147 (Churchwell et al. 1992).
The detectability of any nonthermal radio emission  is 
affected by absorption  and  viewing geometry. As a result, 
nonthermal emission is not seen in all WR binaries.

\section{Gamma-Rays}

The formation of a WR$+$BH binary may be accompanied by
a $\gamma$-ray burst (Tutukov \& Cherepashchuk 2003).
The best candidate for a WR$+$BH system to date
is Cyg  X-3 (Schmutz et al. 1996).
Some models predict that  electrons in the CW region
of WR binaries will be accelerated up to $\gamma$-ray energies
by shocks and inverse Compton scattering 
(e.g. Benaglia \& Romero 2003). 
Initial support for this idea comes from the reported
{\it Fermi}-LAT $\gamma$-ray detection 
of the WR binary $\gamma^2$ Vel  by Pshirkov (2016).
Although this report is encouraging, the detection significance
(6$\sigma$) is relatively low and other WR binaries in the 
Pshirkov (2016) sample were undetected by {\it Fermi}.  
Improvements in the sensitivity and angular resolution
of $\gamma$-ray telescopes will be needed to quantify
the properties of WR binaries in this very high energy 
($\sim$1 - 100 GeV) regime.

\section{Conclusions}\label{sec5}

Wolf-Rayet stars exhibit a diverse range of high-energy
phenomena, largely attributed to their
powerful shocked winds.  But details of the 
underlying emission mechanisms are still not 
well-understood. Hot X-ray plasma is prevalent in WN and WO
stars but temperatures and luminosities are not fully
in accord with shock model predictions. Other factors
such as magnetic fields may be involved in WR X-ray production.
Mysteriously, carbon-rich WC stars remain undetected in X-rays.
Nonthermal (synchrotron) radio emission has been traced
to the colliding wind region in WR binaries, but
the mechanism(s) which accelerate electrons  to
relativistic energies is still debated. 
Gamma-ray emission has been reported from the nearest 
WR binary  $\gamma^2$ Vel. Future improvements in
sensitivity and angular resolution will be needed to confirm 
this intriguing discovery and  search for $\gamma$-rays
in more distant WR systems.


\section*{Acknowledgments}
This work was supported by SAO/CXC award GO6-17009X.

\nocite{*}
\bibliography{skinner.bib}%

\end{document}